# EMERGENCE OF INTRINSIC REPRESENTATIONS OF IMAGES BY FEEDFORWARD AND FEEDBACK PROCESSES AND BIOLUMINESCENT PHOTONS IN EARLY RETINOTOPIC AREAS

**[*][1]I. Bókkon**

[1]Doctoral School of Pharmaceutical and Pharmacological Sciences, Semmelweis University, Hungary

**[2]V. Salari**

[2]Institut de Mineralogie et de Physique des Milieux Condenses, Universite Pierre et Marie Curie-Paris 6, CNRS UMR7590
[2]Boite courrier 115, 4 place Jussieu, 75252 Paris cedex 05, France
[2] BPC Signal, 15 rue Vauquelin, 75005 Paris, France

**and**

**[3]J.A. Tuszynski**

[3]Department of Experimental Oncology, Cross Cancer Institute, 11560 University Avenue Edmonton, AB T6G 1Z2, Canada
[3]Department of Physics, University of Alberta, Edmonton, AB Canada

**2010**

[*]Corresponding author: Bókkon I.
Corresponding author's Email: bokkoni@yahoo.com
Corresponding author's Address: H-1238 Budapest, Lang E. 68. Hungary
Corresponding author's Phone: +36 20 570 6296
Corresponding author's Fax: + 36 1 217-0914

**Abstract** Recently, we put forwarded a redox molecular hypothesis involving the natural biophysical substrate of visual perception and imagery. Here, we explicitly propose that the feedback and feedforward iterative operation processes can be interpreted in terms of a homunculus looking at the biophysical picture in our brain during visual imagery. We further propose that the brain can use both picture-like and language-like representation processes. In our interpretation, visualization (imagery) is a special kind of representation i.e., visual imagery requires a peculiar inherent biophysical (picture-like) mechanism. We also conjecture that the evolution of higher levels of complexity made the biophysical picture representation of the external visual world possible by controlled redox and bioluminescent nonlinear (iterative) biochemical reactions in the V1 and V2 areas during visual imagery. Our proposal deals only with the primary level of visual representation (i.e. perceived "scene").

## 1. Introduction

Recently, a redox molecular hypothesis has been put forward regarding the natural biophysical substrate of visual perception and imagery (see Fig. 1) [6,12]. This novel biophysical hypothesis not only revived Kosslyn's depictive assumption [37] and the *homunculus*, but has also argued that biophysical pictures can emerge in retinotopic visual regions.

Here, we present an iterative model of the *homunculus.* Namely, we suggest that during visual imagery, iterative feedforward and feedback processes can be interpreted in terms of a *homunculus* ("*little man*") looking at the biophysical picture-representation. However, in our hypothesis there is no picture (as 'we see' one on TV) in our brain; rather the biophysical picture is represented by biophotonic signals (see Fig. 1 and section 2 about biophotonic signals) in our brain. There is a real possibility that biophysical pictures are part of the re-entrant feedforward and feedback processes, and they are not separate from each other because of the re-entry [21]. Thus, a separate *homunculus* looking at biophotonic representations can be a misleading concept, because it is a matching process [60,64,66,46]. The matching element is both in physical and mental aspects of feedforward and feedback signals. However, we can render the visual *homunculus* and its mind's eye by showing that it may be reduced to a set of non-linear biophysical iterative processes.

We emphasize that, here, we describe the emerged biophysical pictures and not the conscious interpretation of the emerged biophysical pictures, because interpretation involves consciousness, a language-dependent phenomenon. For example, our hypothesis involving the biophysical picture representation in retinotopic areas probably acts both in primates and in humans but in primates it occurs without conscious (*language*) interpretation. Hence, we do not need to deal with the hard problem of consciousness [16] and do not try to clarify the meaning of consciousness (*there are numerous diverse meanings attributed to the term 'consciousness' in the literature*) [65].

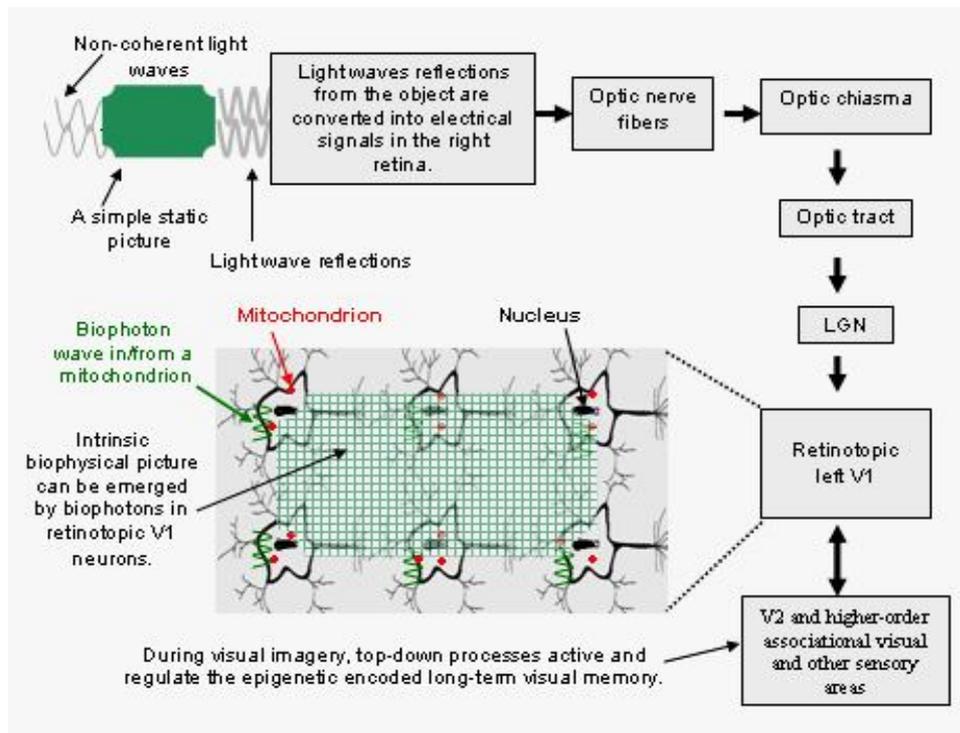

**Fig. 1. Schematic illustration of the biophysical picture representation hypothesis of visual perception and imagery** (Bókkon, 2009; Bókkon and D'Angiulli, 2009). Light waves from a picture are converted into electrical signals in the retina. Next, retinotopic electrical signals are conveyed to V1 and converted into regulated biophotons by mitochondrial redox processes in V1 neurons. Namely, spike-related retinotopic electrical signals - *along classical axonal-dendritic pathways* - produce synchronized bioluminescent biophoton signals by redox processes within retinotopic V1 neurons. Small clusters of visual neurons act as "visual pixels" appropriate to the topological distribution of photon stimuli on the retina. Therefore, we may obtain an intrinsic computational biophysical picture of an object created by biophotons in retinotopic V1. Long-term visual memories are not stored as biophysical pictures but as compressed (epigenetic) codes. We can identify objects because the same compression processes are achieved every time we see an object, and thus what is stored in long-term visual memory will match with what is produced when we see the object again.

During visual imagery, top-down processes regulate the epigenetic encoded long-term visual information. Then, according to retrieved epigenetic information, synchronized neurons generate dynamic patterns of biophotons via redox reactions. Finally, biophotons within synchronized millions of neurons (Bókkon et al., 2010) can produce biophysical pictures in retinotopic and mitochondrial rich visual neurons. We emphasize here that electrical signals are transmitted between neurons but biophotons are produced within retinotopic visual neurons.

## 2. Functional free radicals and regulated ultraweak photon generation in cells and neurons

Recent findings have provided evidence that ROS (reactive oxygen species) and RNS (reactive nitrogen species) as well as their derivatives act as essential regulated signals in biological systems. Namely, reactive species have been identified as second messengers in cells which play essential roles in cell receptor signaling and post-translation modification of signaling molecules, as well as in gene expression, apoptosis, cell growth, cell adhesion, enzymatic functions, $Ca^{2+}$ and redox homeostasis, as well as in numerous other processes [15,20,22,45,47,52,61]. ROS, RNS and their derivatives also act as signaling molecules in cerebral circulation and are indispensable in molecular signal processes, synaptic plasticity, and memory formation under physiological circumstances [5,28,30,32,41].

Ultraweak spontaneous photons (*also called biophotons*) are constantly emitted by living systems at a cell level without any external excitation [17,18,19,33,34,49,58,63,70]. The source of biophotons is due to the diverse biochemical reactions, principally bioluminescent radical reactions of ROS and RNS and the cessation of excited states. The key source of biophotons derives from oxidative metabolism of mitochondria and lipid peroxidation [42,59]. Neural cells also continuously produce biophotons during their ordinary metabolism [27,29]. *In vivo* intensity of biophoton emission from a rat's brain is correlated with cerebral energy metabolism, EEG activity, cerebral blood flow, and oxidative stress [35,36]. In addition, recently Sun et al. [57] revealed that ultraweak bioluminescent photons can conduct along the neural fibers and can be considered a means of neural communication. It has been suggested that biophotonic and bioelectronic activities are not independent biological events in the nervous system, and their synergistic action may play an important role in neural signal transductions.

Recently, Wang et al [68] presented the first experimental proof of the existence of spontaneous ultraweak photon emission and visible light induced delayed ultraweak photon emission from *in vitro* freshly isolated rat's whole eye, lens, vitreous humor and retina. The experiments of Wang et al. [68] indicate that induced photon emission can exist within the eye, and suggest that retinal phosphenes result from excess bioluminescent photons, and the brain interprets these retinal bioluminescent photons as if they originated in the external world. In addition, Bókkon and Vimal [11] pointed out that both retinal phosphenes and the discrete dark noise of rods can be due to the natural redox related (free radical) bioluminescent photons in the retina. Because retinal and cortical phosphenes can have similar mechanisms, if it can be demonstrated that perception of cortical induced phosphenes is due to bioluminescent photons [8,37], intrinsic regulated biophotons in early retinotopic visual system can be seen to serve as a natural biophysical substrate of visual perception and imagery.

Because the production of ROS, RNS and their derivatives is not a random process, but rather a precise mechanism used in cellular signaling pathways, the biophoton production can also be a regulated process. It is worth mentioning that biophoton intensity can be considerably higher inside cells and neurons than that expected from biophoton measurements, which are usually carried out at a distance of several centimeters away from the cells. Our recently published calculations [9] suggest that the real biophoton intensity in retinotopic neurons may be sufficient for creating biophysical picture representation of a single-object image during visual perception. It is entirely plausible that living cells retain their biophotons within the cellular environment for use in signal processing.

Based on the above mentioned functional roles of free radicals and regulated ultraweak biophoton generation in cells and neurons, Bókkon [9] and Bókkon and D'Angiulli [5] put forward a redox molecular hypothesis regarding the natural biophysical substrate of visual perception and visual imagery (see Fig.1). It states that retinotopic electrical signals (*spike-related electrical signals along classical axonal-dendritic pathways*) can be converted into regulated biophoton signals by redox processes that make it possible to produce biophysical picture representation in retinotopically organized mitochondrial cytochrome oxidase-rich visual areas during visual imagery and visual perception (see Fig.1).

## 3. Depictive representation

In cognitive science the long-standing imagery debate involves two rival theories, namely Kosslyn's pictorial theory [37,38] and Pylyshyn's tacit knowledge explanation [48].

According to Pylyshyn, activation of early visual parts is epiphenomenal during visual mental imagery [48]. In addition, mental imagery is explained by language-like representation and can be reduced to tacit knowledge. In particular, we represent objects more abstractly in a propositional format, rather than analogic (*or depictive*) format as speculated by pictorial presumption.

In our interpretation, visualization (*imagery*) is a special kind of representation i.e., visual imagery requires peculiar inherent biophysical processes. Nevertheless, there is growing evidence that visual perception and visual mental imagery share common (*or similar*) neural substrates in the brain. The visual mental imagery abilities require the integrity of brain areas related to vision. The role of the striate cortex (*primary visual area, V1*) in visual mental imagery has been amply demonstrated [31]. In particular, there is evidence that both perception and imagery induce activation in retinotopically organized striate and extrastriate cortex [14,24,55,56]. It is possible that neural correlates of visual perception and imagery are not as strict as was previously assumed [3], but this does not mean that visual perception and imagery could not share very similar neural substrates. However, our brain is not a computer that works by very strict geometrical and algorithmic processes.

## 4. Retinotopic V1 and V2 can represent the principal submodalities of vision such as colour, form, motion and depth

In primates, the main pathway serving visual perception goes from the retina via the lateral geniculate nucleus to V1. One of the most persuasive examples of columnar structure is provided by the distribution of mitochondrial cytochrome oxidase in the primary visual cortex. The V1 and V2 are comprised of regions of various cytochrome oxidase (CO) activities, which can subserve different functions. In V1, layers 2 and 3 are composed of CO-dense patches (blobs) and surrounding regions (*interblobs*) [69]. V2 is composed of alternating thin and thick CO-dense stripes and the pale interstripe regions between them. During visual perception, the high activity of cytochrome oxidase is associated with high mitochondrial activity.

To understand the basic circuitry of vision, it is crucial to know how the projections between V1 and V2 are organized [53]. From V1, most signals are conveyed to the V2 area before distribution to higher cortical areas. Visual areas beyond V1 and V2 have greater specializations for processing different attributes of the visual scene such as colour, form, and motion. The V1 and V2 areas have a rather disordered topographic map of the retina and hence are said to be topographically (*retinotopically*) well organized. There are several further visual areas beyond V1 and V2 in what is known as the prestriate cortex, and they have larger receptive fields and cruder topographic organizations. There is growing evidence that different CO compartments in V1 and V2 are connected in parallel and the projection from V1 cytochrome oxidase blobs (*or patches*) to V2 thin stripes is responsible for colour. However, V1 and V2 can represent all the principal submodalities of vision such as colour, form, motion, and depth [2]. V1 sends most of its cortical output to V2 and in return receives a strong feedback projection. V1 and V2 contain similarly scaled retinotopic maps of the visual field and both have comparable surface areas [53].

## 5. Fast feedback and feedforward conduction velocities between V1 and V2

V1 area contains about 11,000 feedback neurons and V2 area includes about 14,000 feedforward neurons [51]. It is believed that the action of cortical feedback connections is slow, whereas feedforward connections can carry a rapid drive to their target neurons. According to recent results [26], the effects of feedback connections can be delayed by less than 10 ms with respect to the beginning of the responses of neurons in low-order visual parts. That is, there is an especially rapid effect of feedback connections on the visual responses of neurons in lower order areas [25]. These feedback and feedforward processes between V1 and V2 have similar fast conduction velocities (*around 3.5 m/s*).

## 6. V1- V2 and biophysical retinotopic representation

Since the retina employs depictive representation the question arises why LGN makes it again, which is also retinotopically organized. Why would V1 or V2 be also retinotopically well organized instead of discarding this information? The electric coding of the visual photic signal should not require a multiple retinotopic neural coding. In addition, Slotnick [54] has recently reported evidence for the existence of retinotopic areas in frontal and parietal cortex during spatial attention and working memory. These retinotopic regions interact during retrieval of spatial information. This multiple retinotopic organization should have to perform some special function during visual perception and imagery.

Some summarized important characteristics of V1 and V2:
**i.** The V1 and V2 areas are well organized retinotopically and preserve the local spatial geometry of the retina, so patterns of activation in them depict shape [38,67]. **ii.** V1 and V2 have comparable surface areas [53]. **iii.** V1 mitochondrial-rich cytochrome oxidase patch columns project onto V2 mitochondrial-rich cytochrome oxidase thin stripes [53]. **iv.** There are about 11,000 feedback neurons in V2 and 14,000 feedforward neurons in V1 [51]. **v.** There are very rapid feedforward and feedback processes between V1 and V2 with fast conduction velocities (around 3.5 m/s) [25,26]. **vi.** A map of V2 approximates a mirror image of the V1 [71]. **vii.** There is growing evidence that both visual perception and imagery induce activation in retinotopically organized striate and extrastriate cortex and share very similar neural substrates in the brain [14,24,55,56]. **viii.** There may be iterative non-linear computations taking place during vision [4,40,44].

According to Bókkon [12] and Bókkon and D'Angiulli [6], during visual imagery, top-down processes activate and regulate the epigenetic[*] encoded long-term visual information. This epigenetic information can be retrieved by neurocellular redox and radical processes during visual imagery. Then, the retrieved long-term visual redox information can be converted into regulated bioluminescent photons by redox (free radical) processes that make it possible to generate biophysical pictures in retinotopic areas during visual imagery (see Fig.1). We should also mention that early retinotopic areas are essential for visual apperception. Namely, according to Bókkon and Vimal [10] "synchronized (*coupled*) mitochondrial processes, the duration of visual representation (*that is also needed to the feedforward* and feedback *iterative/recursive processes*), and specific mitochondrial-rich retinotopic structures in V1 are elementary conditions for the emergence of explicit conscious visual perception".

Since imagery can induce activation in retinotopically organized V1 and V2, we can have two similar, biophysical retinotopic representations with a very short delay time in V1 compared to V2. Rapid conduction velocities along feedforward and feedback axons between V1 and V2 can make fast non-linear iterative redox computation possible and can give rise to the feeling that a visual *homunculus* is looking at the biophysical picture in our brain during visual imagery. We can explicitly show that the visual *homunculus* can be reduced to a set of biophysical iterative processes. We emphasize again that a separate *homunculus* looking at a biophotonic representation can be puzzling, because it represent a matching process that carries out the *homunculus'* job.

If we take the V1 and V2 listed characteristics into consideration, this may show the feasibility of the above described simple process. In other words, the emerged topographic biophysical pictures in V1 and V2 can "see each other" with a very short delay time. However, the interpretation of emerged mirror topographic biophysical pictures can be achieved by higher-order associational visual and other sensory areas during visual imagery. We should consider that emergence of an iterative biophysical picture by biophotons in V1/V2 and the semantic interpretations of the emerged biophysical picture are two different but tightly connected issues. The first is a biophysical picture-generating process (*picture-like*) in early retinotopic areas, while the second is a language-like semantic interpretation process.

Previously, we suggested that dynamic series of picture-representations can carry unambiguous meaning of words [7,12,13]. The human memory can operate through inherent dynamic picture-representations and we link these biophysical pictures to each other during language learning processes. During learning processes, picture-like and language-like representations become quasi-independent neural processes. It means that our brain can use both picture-like and language-like representation processes. The language-like processes can become the basis of abstract thinking, interpersonal communication, etc., while the picture-like biophysical representation processes can guarantee computational geometric imaginary events. For example, they can envision or compose and design geometric things, etc. Language-like and picture-like processes are tightly connected, which can induce each other's representations. The important implication is that long-term information storage of the language-like and picture-like representations can be linked and encoded by non-linear neuroredox processes at an epigenetic level.

**\*Footnote**

The latest studies suggest that epigenetic modulation of the genome (*i.e., the regulation of chromatin structure through direct methylation of DNA or post-translational modification of histone proteins, including methylation, acetylation, and phosphorylation*) is a necessary component for the formation of neuronal plasticity, associative learning and long-term memory [1,23,50]. Chromatin structure itself can represent a "memory" and allow for temporal integration of spaced signals or metaplasticity of synapses [39]. The epigenetic model, which states that the long-term memory is stored at the level of modified DNA molecules, has obtained some recognition, and appears to hold promise.

## 7. Mind's eye, mind's ear, mind's skin, and so forth

The concept of a *homunculus* (Latin for "little man", sometimes spelled "*homonculus*") is frequently used to demonstrate the functioning of a mental system. In the scientific sense of an unknowable prime actor, it can be viewed as an entity. Who looks at the images in the brain? If we presume that this is a *homunculus* who does it, our visual imagery is associated with the occurrence of seeing with the mind's eye. Nevertheless, auditory mental imagery is also accompanied by the experience of hearing with the mind's ear or tactile imagery is accompanied by the experience of feeling with the mind's skin, and so forth. Although it is controversial to assume that the brain perfroms iterative processes during vision [4,40,44], the visual *homunculus*, which is a matching process, may be achieved by iterative processes.

## 8. Emergence of biophysical pictures by simple iterative processes and biophotons between V1 and V2

According to recent findings [25] the effects of feedback connections are delayed by less than 10 ms with respect to the beginning of the responses of neurons in low-order visual areas. The feedback and feedforward connections between V1 and V2 have comparable fast electric conduction velocities (around 3.5 m/s). We assume that the neurons in layers V1 and V2 are similar to lattices composed of pixels in which each neuron is equivalent to a pixel. A static electric field is generated by a static charged particle. Both an electric field and a magnetic field are generated if a charged particle moves at a constant velocity. Electromagnetic radiation is produced when a charged particle is accelerated. If the frequency of an oscillating charge is high and approaches the optical part of the EMF spectrum, it generates photons [72]. The generation of photons is usually interpreted as a process where a charged particle "drops" from a higher energy (excited) state to a lower energy (ground) state ($h c/\lambda = E_2 - E_1$), where $h$ is Planck's constant, $c$ is the speed of light, $\lambda$ is wavelength of the photon, $E_2$ is the energy of the excited state and $E_1$ the energy of the ground state [73]. Various cell functions are associated with moving charges in cellular compartments and can generate electromagnetic radiation [72]. According to the previous sections when information (i.e. in the form of electric charges) reaches to the neurons (i.e. pixels) they can produce biophotons.

To explain simply how a biophysical picture emerges in V1 and V2 via iterative processes, we demonstrate it here using the language of mathematics. For simplicity we consider that V1 and V2 are represented using matrices $V_1$ and $V_2$, respectively. The elements of these matrices are equivalent with the pixels of V1 and V2, which are visual neurons. Each visual neuron can produce intrinsic biophotons. We emphasize here that electrical signals are transmitted between neurons but biophotons are produced within retinotopic visual neurons. Namely, each visual neuron can produce intrinsic biophotons after receiving electrical information. Biophotons can be produced with different frequencies (i.e. different wavelengths) [62]. Therefore, in general each visual neuron can be represented by an element $v_{ij}^k$ $(k = 1,...,n)$, which is the *ij*-th element of matrix *V*. Each element produces *k* biophotons in which each biophoton has a special wavelength. The number *k* indicates the number of biophotons in each stage.

V1 and V2 receive two electrical information inputs: one from LGN, which is external information from the eye, and the other, is internal information, which is transformed between V1 and V2 and vice versa. We represent the input information as matrices $E_i$ which have the same dimensions as matrices $M_1$ and $M_2$. The iterative process for seeing an image can be explained using the lattices shown below (see Fig. 2a, 2b, 2c, 2d). The lattices on the left are V1 and those on the right are V2. The mathematical illustrations are given in terms of information.

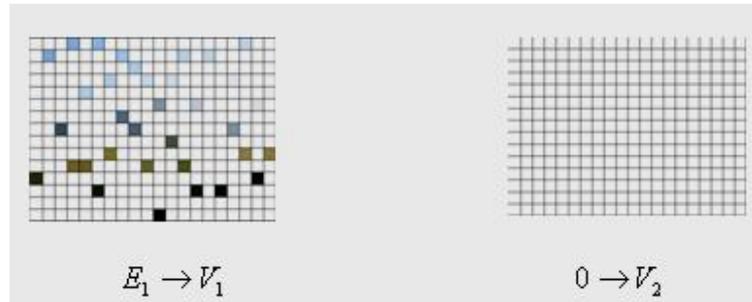

**Fig. 2a. In the first stage the layer V1 receives information $E_1$ from LGN and layer V2 has not received any information yet (i.e. zero).**

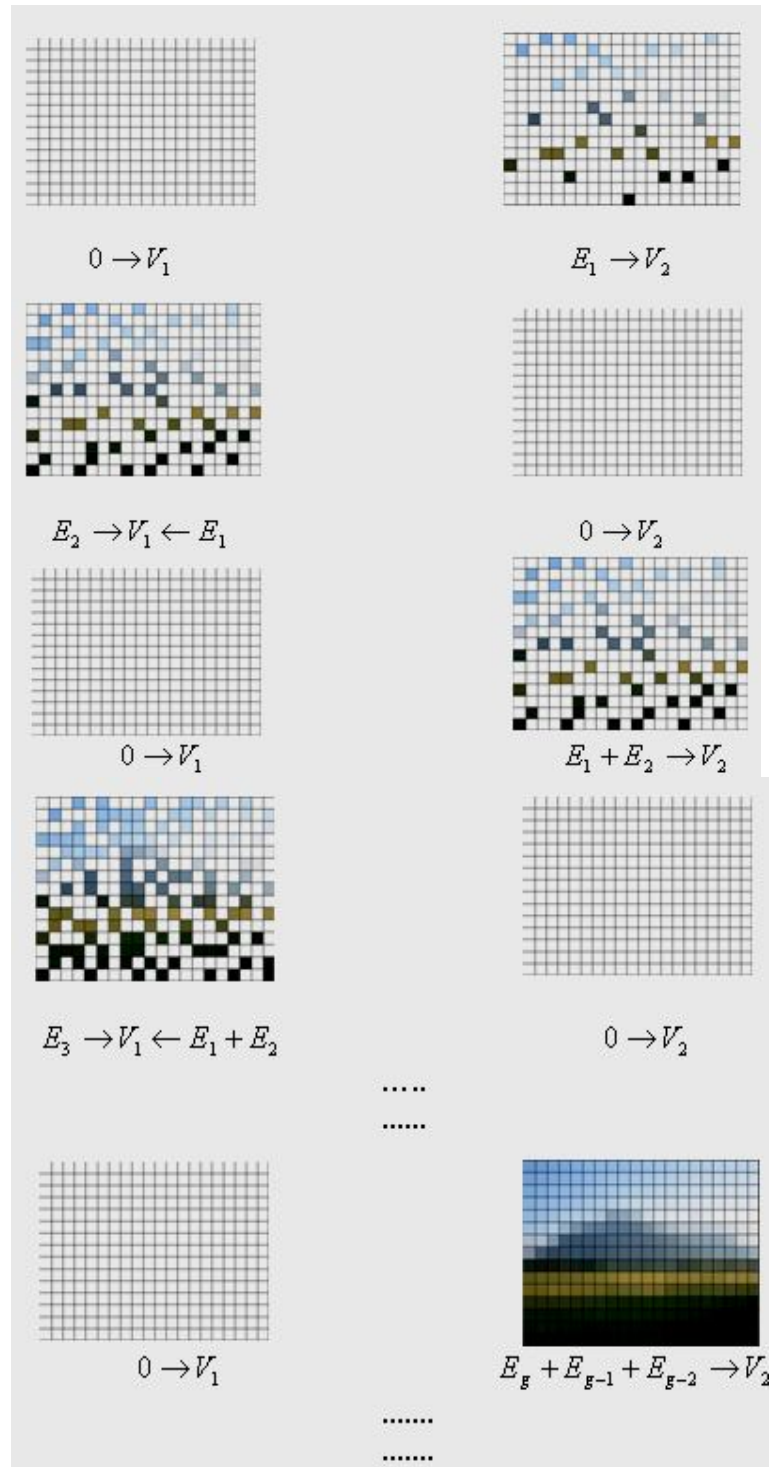

**Fig. 2b. In the iterative process the electrical information is transferred from V1 to V2 and vice versa. The process will be repeated again until the whole neurons in each layer becomes activated for production of a complete biophysical picture by biophotons. When information is activated in V1 there is no active information in V2 any more and vice versa.**

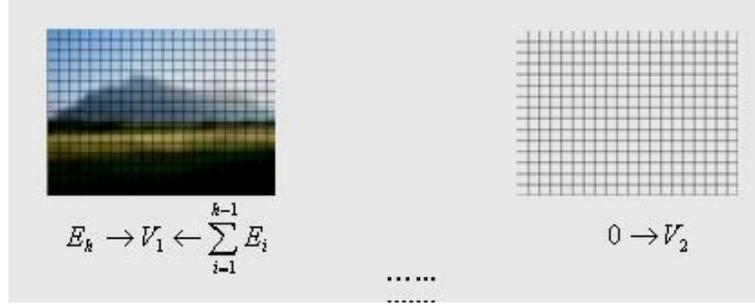

**Fig. 2c. At this stage the iterative process happened k times between V1 and V2. The information $E_k$ from LGN and the information $\sum_{i=1}^{k-1} E_i$ from V2 arrive at V1 simultaneously.**

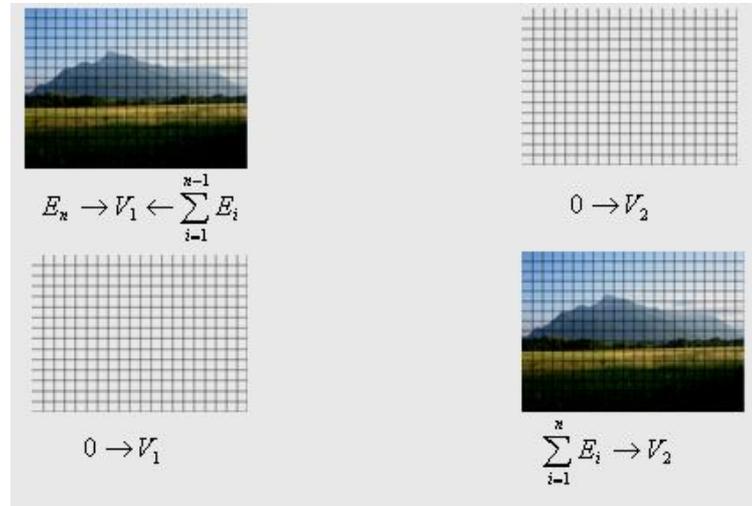

**Fig. 2d. The final stage of iterative process. After n-1 iterations, information from LGN and V2 enters into V1, and then the whole information in V1 (i.e. $E_n + \sum_{i=1}^{n-1} E_i = \sum_{i=1}^{n} E_i$ ) is completely transmitted to V2**

We define $w$ as the total number of neurons, and $s_i$ ( $s_i < w$ ) is the number of neurons in stage $i$ which are activated (i.e which produce intrinsic biophotons within neurons) for perception of an object in each lattice of V1 or V2, and $p$ is the total number of biophotons in the last stage of iteration to emerge as a biophotonic representation of a conscious event. It means that $p$ is the number of internal biophotons involved in the perception of an object.

It is clear that for $s_i$ $(i=1,...,n)$ we have $s_1 < s_2 < s_3 < ... < s_n$, where $s_1$ is a low-resolution state and $s_n$ (i.e. final stage) is a high-resolution state. As presented above, $v_{ij}^k$ is the element of matrix $V_i$ $(i=1,2)$, which can produce biophotons up to $k$ biophotons. According to our arguments in previous sections the resolution of matrices becomes two times higher in each stage $m$ relative to $m-1$, where $m$ is an arbitrary number and $m < n$. According to the iterative process, it seems reasonable to assume that the transfer of electrical information from LGN to V1 activates the same number of neurons in V1 in each step. Therefore, the simultaneously received information in V1 from both LGN and V2 would make the V1 cells twice more active in each iteration step.

Thus we have

$$s_2 k_2 = 2 s_1 k_1$$
$$s_3 k_3 = 2 s_2 k_2 = 2^2 s_1 k_1$$
$$\ldots..$$
$$s_n k_n = 2 s_{n-1} k_{n-1} = 2^{n-1} s_1 k_1$$

Thus in the final stage the visual cortex produces $p = s_n k_n$ biophotons which is the stage of a consciousness moment. According to the calculations of Bókkon et al. [9], the number of biophotons produced within neurons of the visual cortex when seeing a single object is on the order of $10^8$-$10^9$, so if we assume that $p=10^8$ we have

$$p = s_n k_n = 2^{n-1} s_1 k_1$$

If we let $p_1 = s_1 k_1$, then for $p = 10^8$ we have

$$n = \frac{8 - \log p_1}{\log 2} + 1$$

This indicates that the number of iterations depends on the number of the first stage of biophoton production in the iterative process between V1 and V2. Figure 3a shows that the number of iterations $n$ for this state is in the range $25 < n < 35$.
Then, for $p=10^9$ we have

$$n = \frac{9 - \log p_1}{\log 2} + 1$$

According to Figure 3b, the number of iterations $n$ is in the range of $29 < n < 40$.

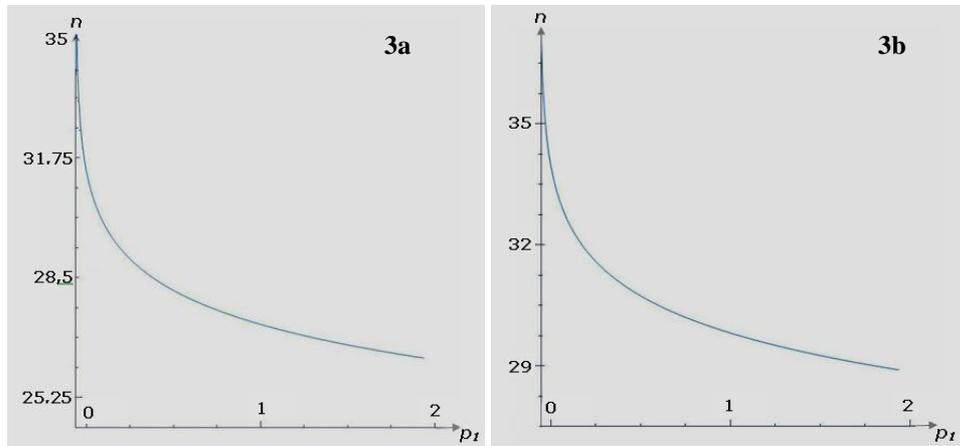

**Fig. 3.** The number of iterative processes vs $p_1$ for (3a) $10^8$ biophotns and (3b) $10^9$ biophotons for the final stage of iteration in V1 and V2.

In general, according to the above diagrams we have at least 25 iterative processes between V1 and V2. Also, if we consider that a conscious process takes about 300-400 milliseconds and according to [26] the delay time for each iteration is around 10 milliseconds (*or less*), then the number of iterations is about 30-40 which is very close to our estimate given in the above calculation.

During visual imagery similar iterative processes can be carried out in the same way as in visual perception, although signals originate from the long-term visual information. Namely, the top-down processes trigger and regulate the epigenetically encoded long-term visual information during visual imagery. Then, according to retrieved epigenetic information, mitochondrial networks in synchronized neurons generate dynamic patterns of bioluminescent biophotons via redox reactions. Finally, synchronized dynamic patterns of biophotons can produce biophysical pictures (depictive representation) in retinotopic visual neurons of V1 and V2 via iterative processes. Figure 4 is a schematic drawing of biophysical visual imagery with the help of iterative processes.

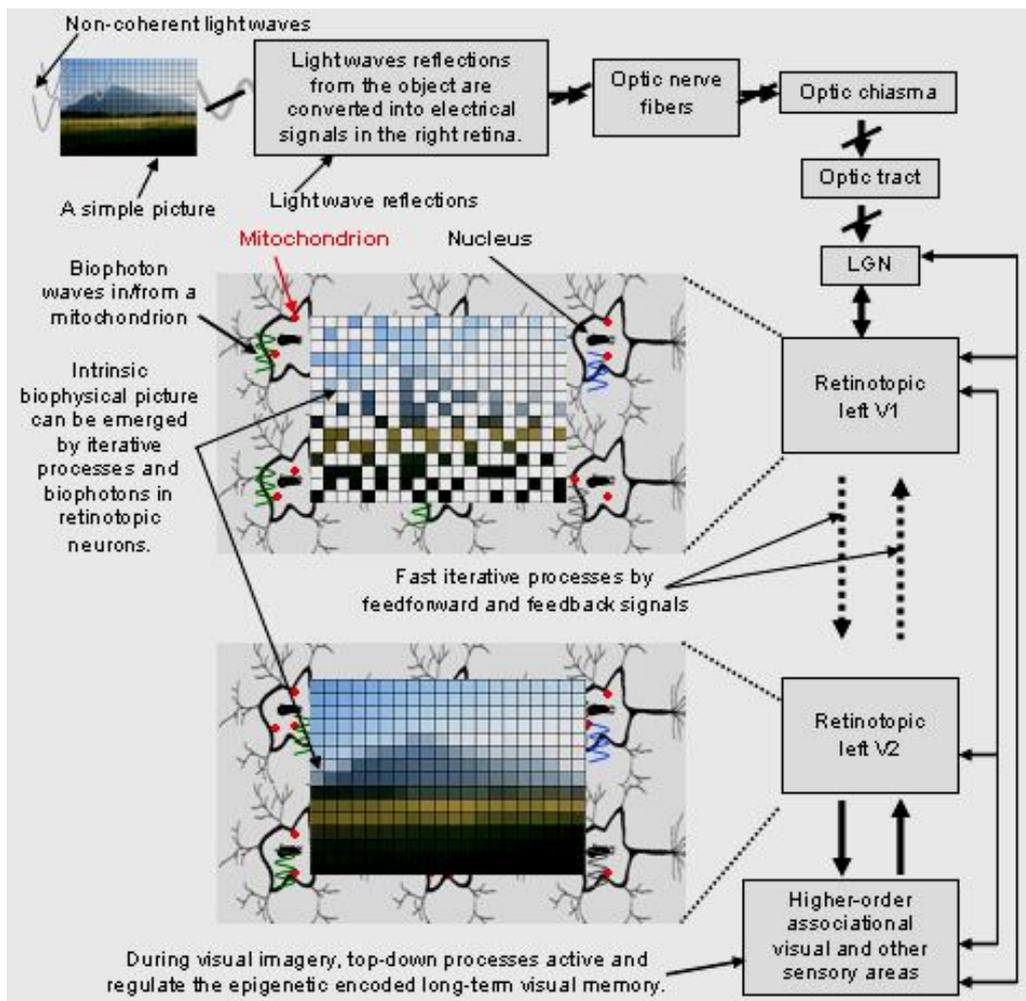

**Fig. 4. Schematic depiction of visual imagery by feedforward and feedback and biophotons in early retinotopic areas.** During visual imagery, top-down processes activate and regulate the epigenetic encoded long-term visual memory. Next, according to retrieved long-term information, mitochondrial networks within synchronized neurons produce dynamic patterns of biophotons via redox reactions. These dynamic patterns of biophotons can produce biophysical pictures (depictive representation) in retinotopic and mitochondrial rich visual neurons by iterative processes. As a result, we could retrieve what we thought we would have seen or done in the analogous perceptual situation during visual imagery.

## 9. Summary

In this paper, we have proposed a theoretical model involving a biophysical picture-representation without *homunculus* during visual imagery. We do not claim to have explained the enigma of consciousness, but our goal was to show that the somewhat mysterious *homunculus* phenomenon may be elucidated with the help of retinotopic representation, rapid feedforward and feedback connections (*between V1 and V2*), and non-linear iterative processes during visual imagery. We also proposed that emergence of an iterative biophysical picture-representation in retinotopic V1/V2 and the semantic interpretation of an emerged biophysical picture are two different things, although they may be tightly connected. The first is a biophysical picture-representation generating process (*picture-like*) while the second is a language-like semantic interpretation process. However, they can induce each other's representations.

The human memory can operate through intrinsic dynamic pictures and we link these picture-representations to each other during language learning processes. During language learning processes, development of picture-like and language-like systems becomes a quasi-independent neural process. An important implication of this hypothesis is that long-term information storage of the language-like and picture-like representations can be encoded by non-linear epigenetic redox processes. The evolutionary advantage of the biophysical picture representation is that it makes possible, for example, for us to imagine events, compose and design objects, etc.

However, if it can be proved that perception of cortical induced phosphene lights is due to biophotons; intrinsic regulated biophotons in the brain may serve as a natural biophysical (redox molecular) substrate of visual perception and imagery. In other words, intrinsic biophysical visual virtual reality may emerge from feedback and feedforward iterative operation processes and biophotons in early retinotopic V1 and V2 areas. Kosslyn's reality simulation principle [38] states that mental imagery mimics the corresponding events in the world. However, our concept of intrinsic biophysical visual virtual reality (*by iterative processes*) in retinotopic areas may be nothing else than a possible biophysical basis of the reality simulation principle in the case of visual imagery.

**Acknowledgments**
Bókkon I. gratefully acknowledges support of this work by the BioLabor Hungary, www.biolabor.org, His URL: http://bokkon-brain-imagery.5mp.eu. The authors thank the anonymous reviewers for constructive suggestions which were very helpful in improving of our paper. The authors thank Jonathan Mane (PhD) for his helps to prepare the manuscript.

**References**

[1] Arshavsky YI. „The seven sins" of the Hebbian synapse: can the hypothesis of synaptic plasticity explain long-term memory consolidation? *Prog Neurobiol* **80**:99–113, 2006.
[2] Bartels A, Zeki S, The theory of multistage integration in the visual brain, *Proc Biol Sci* **265**:2327–2332, 1998.
[3] Bartolomeo P, The neural correlates of visual mental imagery: an ongoing debate, *Cortex* **44**:107–108, 2008.
[4] Beck C, Neumann H, Interactions of motion and form in visual cortex - A neural model, *J Physiol Paris* **104**:61–70, 2010.
[5] Bókkon I, Antal I, Schizophrenia: redox regulation and volume transmission, *Curr Neuropharmacol* 2010 In press.
[6] Bókkon I, D'Angiulli A, Emergence and transmission of visual awareness through optical coding in the brain: A redox molecular hypothesis on visual mental imagery, *Bioscience Hypotheses* **2**:226–232, 2009.
[7] Bókkon I, Dreams and Neuroholography: An interdisciplinary interpretation of development of homeotherm state in evolution, *Sleep Hyp* **7**:61–76, 2005.
[8] Bókkon I, Phosphene phenomenon: a new concept, *BioSystems* **92:**168–174, 2008.
[9] Bókkon I, Salari V, Tuszynski J, Antal I, Estimation of the number of biophotons involved in the visual perception of a single-object image: Biophoton intensity can be considerably higher inside cells than outside, *J Photochem Photobiol B* **100**: 160–166, 2010.
[10] Bókkon I, Vimal RLP. Implications on visual apperception: energy, duration, structure and synchronization, *BioSystems* **101:**1–9, 2010.
[11] Bókkon I, Vimal RLP. Retinal phosphenes and discrete dark noises in rods: A new biophysical framework. *J Photochem Photobiol B* **96**:255–259, 2009.
[12] Bókkon I. Visual perception and imagery: a new hypothesis, *BioSystems* **96**:178–184, 2009.
[13] Bókkon I., Dream pictures, neuroholography and the laws of physics. *J Sleep Res* **15** (Suppl. 1):187, 2006.
[14] Borst G, Kosslyn SM, Visual mental imagery and visual perception: structural equivalence revealed by scanning processes. *Mem Cognit* **36**:849–862, 2008.
[15] Brookes PS, Levonen AL, Shiva S, Sarti P, Darley-Usmar VM, Mitochondria: regulators of signal transduction by reactive oxygen and nitrogen species. *Free Radic Biol Med* **33**:755–764, 2002.
[16] Chalmers DJ, Facing Up to the Problem of Consciousness, *J Consciousness Studies* **2**: 200–219, 1995.
[17] Chang JJ, Physical properties of biophotons and their biological functions, *Indian J Exp Biol* **46**:371–377, 2008.
[18] Cohen S, Popp FA, Biophoton emission of the human body, *J Photochem Photobiol B* **40**:187–189, 1997.
[19] Devaraj B, Scott RQ, Roschger P, Inaba H, Ultraweak light emission from rat liver nuclei, *Photochem Photobiol* **54**:289–293, 1991.
[20] Dröge W, Free radicals in the physiological control of cell function, *Physiol Rev* **82**:47–95, 2002.
[21] Edelman GM. Neural Darwinism: selection and reentrant signaling in higher brain function, *Neuron* **10**:115–125, 1993.

[22]     Feissner RF, Skalska J, Gaum WE, Sheu SS, Crosstalk signaling between mitochondrial Ca2+ and ROS, *Front Biosci* **14**:1197–1218, 2009.

[23]     Feng J, Fouse S, Fan G. Epigenetic regulation of neural gene expression and neuronal function, *Pediatr Res* **61**:58R–63R, 2007.

[24]     Ganis G, Thompson WL, Kosslyn SM, Brain areas underlying visual mental imagery and visual perception: an fMRI study, *Brain Res Cogn Brain Res* **20**:226–241, 2004.

[25]     Girard P, Hupé JM, Bullier J, Feedforward and feedback connections between areas V1 and V2 of the monkey have similar rapid conduction velocities, *J Neurophysiol* **85**:1328–1331, 2001.

[26]     Hupé JM, James AC, Girard P, Lomber SG, Payne BR, Bullier J, Feedback connections act on the early part of the responses in monkey visual cortex, *J Neurophysiol* **85**:134–145, 2001.

[27]     Isojima Y, Isoshima T, Nagai K, Kikuchi K, Nakagawa H, Ultraweak biochemiluminescence detected from rat hippocampal slices, *NeuroReport* **6**:658–660, 1995.

[28]     Kamsler A, Segal M, Hydrogen peroxide modulation of synaptic plasticity, *J Neurosci* **23**:269–276, 2003.

[29]     Kataoka Y, Cui Y, Yamagata A, Niigaki M, Hirohata T, Oishi N, Watanabe Y,. Activity-Dependent Neural Tissue Oxidation Emits Intrinsic Ultraweak Photons, *Biochem Biophys Res Commun* **285**:1007–1011, 2001.

[30]     Kishida KT, Klann E, Sources and targets of reactive oxygen species in synaptic plasticity and memory, *Antioxid Redox Signal* **9**:233–244, 2007.

[31]     Klein I, Dubois J, Mangin JF, Kherif F, Flandin G, Poline JB, Denis M, Kosslyn SM, Le Bihan D, Retinotopic organization of visual mental images as revealed by functional magnetic resonance imaging, *Brain Res Cogn Brain Res* **22**:26–31, 2004.

[32]     Knapp LT, Klann E, Potentiation of hippocampal synaptic transmission by superoxide requires the oxidative activation of protein kinase C, *J Neurosci* **22**:674–683, 2002.

[33]     Kobayashi M, Inaba H, Photon statistics and correlation analysis of ultraweak light originating from living organisms for extraction of biological information, *Appl Opt* **39**:183–192, 2000.

[34]     Kobayashi M, Kikuchi D, Okamura H, Imaging of ultraweak spontaneous photon emission from human body displaying diurnal rhythm, *PLoS One* **4**:e6256, 2009.

[35]     Kobayashi M, Takeda M, Ito KI, Kato H, Inaba H, Two-dimensional photon counting imaging and spatiotemporal characterization of ultraweak photon emission from a rat's brain in vivo, *J Neurosci Methods* **93**:163–168, 1999.

[36]     Kobayashi M, Takeda M, Sato T, Yamazaki Y, Kaneko K, Ito K, Kato H, Inaba H, In vivo imaging of spontaneous ultraweak photon emission from a rat's brain correlated with cerebral energy metabolism and oxidative stress, *Neurosci Res* **34**:103–113, 1999.

[37]     Kosslyn SM, *Image and brain: The resolution of the imagery debate*, MIT Press, 1994.

[38]     Kosslyn SM, Remembering Images, in Gluck MA, Anderson JR, Kosslyn SM (eds.), *Memory and Mind: A Festschrift for Gordon H. Bower*, New Jersey, Lawrence Erlbaum Associates, 2007.

[39]     Levenson JM, Sweatt JD. Epigenetic mechanisms in memory formation, *Nat Rev Neurosci* **6**:108–118, 2005.

[40]     Mahani AS, Wessel R, Iterative cooperation between parallel pathways for object and background motion. *Biol Cybern* **95**:393–400, 2006.

[41] Massaad CA, Klann E, Reactive oxygen species in the regulation of synaptic plasticity and memory, *Antioxid Redox Signal* 2010 Jul 22. [Epub ahead of print].

[42] Nakano M, Low-level chemiluminescence during lipid peroxidations and enzymatic reactions, *J Biolumin Chemilum* **4**:231–240, 2005.

[43] Narici L, De Martino A, Brunetti V, Rinaldi A, Sannita WG, Paci M, Radicals excess in the retina: A model for light flashes in space, *Rad Meas* **44**:203–205, 2009.

[44] Okada M, Nishina S, Kawato M, The neural computation of the aperture problem: an iterative process, *NeuroReport* **14**:1767–1771, 2003.

[45] Palumaa P, Biological redox switches, *Antioxid Redox Signal* **11**:981–983, 2009.

[46] Perlovsky LI. "Vague-to-crisp" neural mechanism of perception, *IEEE Trans Neural Netw* **20**:1363–1367, 2009.

[47] Pourova J, Kottova M, Voprsalova M, Pour M, Reactive oxygen and nitrogen species in normal physiological processes, *Acta Physiol (Oxf)* **198**:15–35, 2010.

[48] Pylyshyn ZW, *Seeing and visualizing: It's not what you think*, MIT Press, 2003.

[49] Quickenden TI, Que Hee SS, Weak luminescence from the yeast Sachharomyces-Cervisiae, *Biochem Biophys Res Commun* **60**:764–770, 1974.

[50] Reul JM, Chandramohan Y, Epigenetic mechanisms in stress-related memory formation, *Psychoneuroendocrinology* 3**2**(Suppl. 1):S21–S25, 2007.

[51] Rockland KS, Elements of cortical architecture: hierarchy revisited, in Rockland KS, Kaas JH, Peters A (eds.), *Cerebral cortex: Extrastriate cortex in primates,* New York, Plenum Press, 1997.

[52] Shiva S, Moellering D, Ramachandran A, Levonen AL, Landar A, Venkatraman A, Ceaser E, Ulasova E, Crawford JH, Brookes PS, Patel RP, Darley-Usmar VM, Redox signalling: from nitric oxide to oxidized lipids, *Biochem Soc Symp* **71**:107–120, 2004.

[53] Sincich LC, Jocson CM, Horton JC, Neurons in V1 patch columns project to V2 thin stripes, *Cereb Cortex* **17**:935–941, 2007.

[54] Slotnick SD, Synchronous retinotopic frontal-temporal activity during long-term memory for spatial location, *Brain Res* **1330**:89–100, 2010.

[55] Slotnick SD, Visual memory and visual perception recruit common neural substrates, *Behav Cogn Neurosci Rev* **3**:207–221, 2004.

[56] Stokes M, Thompson R, Cusack R, Duncan J, Top-down activation of shape-specific population codes in visual cortex during mental imagery, *J Neurosci* **29**:1565–1572, 2009.

[57] Sun Y, Wang Ch, Dai J, Biophotons as neural communication signals demonstrated by in situ biophoton autography, *Photochem Photobiol Sci* **9**:315–322, 2010.

[58] Takeda M, Kobayashi M, Takayama M, Suzuki S, Ishida T, Ohnuki K, Moriya T, Ohuchi N, Biophoton detection as a novel technique for cancer imaging, *Cancer Sci* **95**:656–661, 2004.

[59] Thar R, Kühl M, Propagation of electromagnetic radiation in mitochondria? *J Theor Biol* **230**:261–270, 2004.

[60] Trehub A. Space, self, and the theater of consciousness, *Conscious Cogn* **16**:310–330, 2007.

[61] Valko M, Leibfritz D, Moncol J, Cronin MT, Mazur M, Telser J, Free radicals and antioxidants in normal physiological functions and human disease, *Int J Biochem Cell Biol* **39**:44–84, 2007.

[62] Van Wijk R, Ackerman JM, van Wijk EP. Effects of a color filter used in auriculomedicine on ultraweak photon emission of the human body, *J Altern Complement Med* **12**:955–962, 2006.

[63] Van Wijk R, Van Wijk EP, Bajpai RP, Photocount distribution of photons emitted from three sites of a human body, *J Photochem Photobiol B* **84**:46–55, 2006.

[64] Vimal RLP. Matching and selection of a specific subjective experience: conjugate matching and subjective experience, *J Integr Neurosci* **9**:193-251, 2010.

[65] Vimal RLP. Meanings attributed to the term 'consciousness': an overview, *J Consciousness Studies: Special Issue on Defining consciousness (Ed. Chris Nunn)* **16**:9–27, 2009.

[66] Vimal RLP. Proto-experiences and Subjective Experiences: Classical and Quantum Concepts, *J Integr Neurosci* **7**:49–73, 2008.

[67] Wandell BA, Brewer AA, Dougherty RF, Visual field map clusters in human cortex, *Philos Trans R Soc Lond B Biol Sci* **360**:693–707, 2005.

[68] Wang Ch, Bókkon I, Dai J, Antal I, Spontaneous and visible light-induced ultraweak photon emission from rat eyes, *Brain Res* in press, DOI: 10.1016/j.brainres.2010.10.077, 2010.

[69] Xiao Y, Felleman DJ, Projections from primary visual cortex to cytochrome oxidase thin stripes and interstripes of macaque visual area 2, *PNAS USA* **101**:7147–7151, 2004.

[70] Yoon YZ, Kim J, Lee BC, Kim YU, Lee SK, Soh KS, Changes in ultraweak photon emission and heart rate variability of epinephrine-injected rats, *Gen Physiol Biophys* **24**: 147–159, 2005.

[71] Zeki SM, Simultaneous anatomical demonstration of the representation of the vertical and horizontal meridians in areas V2 and V3 of rhesus monkey visual cortex, *Proc R Soc Lond B Biol Sci* **195**:517–523, 1977.

[72] Cifra M, Fields JZ, Farhadi A, Electromagnetic cellular interactions", *Progress in Biophysics and Molecular Biology*, in press, DOI:10.1016/j.pbiomolbio.2010.07.003, 2010.

[73] Jackson JD, *Classical Electrodynamics,* New York, Wiley, 1975.